\documentclass[pre,pre,twocolumn,floatfix,showpacs]{revtex4}
\usepackage{graphicx}
\usepackage{epsfig}
\usepackage{amssymb,amsmath}
\usepackage{bm}

\def\3dots{\:\raisebox{-0.5ex}{$\stackrel{\textstyle.}{:}$}\:}
\def\beq{\begin{equation}}
\def\eeq{\end{equation}}
\def\bea{\begin{eqnarray}}
\def\eea{\end{eqnarray}}
\DeclareMathOperator{\sech}{sech}
\begin{document}
\title{Mechanics and force transmission in soft composites of rods in elastic gels}
\author{Moumita Das}
\author{F.C.\ MacKintosh}
\affiliation{Department of Physics and Astronomy, Vrije Universiteit, Amsterdam, The Netherlands. \\
 }

\begin{abstract}
We report detailed theoretical investigations of the micro-mechanics and bulk elastic properties of composites consisting of randomly distributed stiff fibers embedded in an elastic matrix in two and three dimensions. 
%It has been known, for instance, that novel elastic behavior can result from the embedding of particles or rods in a matrix. 
Recent experiments published in Physical Review Letters [102, 188303 (2009)] have suggested that the inclusion of stiff microtubules in a softer, nearly incompressible biopolymer matrix can lead to emergent compressibility. 
This can be understood in terms of the enhancement of the compressibility of the composite relative to its shear compliance as a result of  the addition of  stiff rod-like inclusions. 
We show that the Poisson's ratio $\nu$ of such a composite evolves with increasing rod density towards a particular value, or {\em fixed point},  independent of the material properties of the matrix,  so long as it has a finite initial compressibility. 
This fixed point is $\nu=1/4$ in three dimensions and $\nu=1/3$ in two dimensions.
Our results suggest an important role for stiff filaments such as microtubules and stress fibers in cell mechanics. 
At the same time, our work has a wider elasticity context, with potential applications to composite elastic media with a wide separation of scales in stiffness of its constituents such as carbon nanotube-polymer composites, which have been shown to have highly tunable mechanics.
\end{abstract}
\date{\today}
\pacs{87.16.Ka, 62.20.Dc, 82.35.Pq}
 %82.35.Pq Biopolymers, biopolymerization
% 62.20.Dc Elastic constants (in cond-mat section)
 %87.16.Ka Filaments, microtubules, their networks - and supramolecular assemblies
\maketitle

\section{Introduction}

Many materials, natural \cite{alberts,howard} as well as man-made \cite{nanotube1}, that we come across in our daily lives are composite materials, combining multiple components with distinct elastic properties. Notable examples of natural composites are wood, bone, plant and animal cells, while glass or carbon fibers embedded in epoxy resins are two extensively used man-made composites. A hallmark of such materials is that the components they are made up of interact in a highly synergistic manner such that the collective properties are more than merely the sum total of those of the constituents.  These composites are often made up of a soft elastic background, reinforced with stiff fibers.  Because of their aspect ratio, fiber reinforcements have inherent advantages over other geometries allowing, for example, for long range force propagation or enhanced strength along a given direction. By tuning the concentration of the fibers and the relative mechanical properties of the fibers and the matrix, one can modulate the bulk properties of the composite resulting in materials with remarkable properties. Fiber-reinforced composites often form the building block  of structures that require high strength and stiffness as well as low weight, both in materials optimized by nature through evolutionary processes as well as those engineered by humans. 

A prime example of a natural or living composite is the cell cytoskeleton, a composite polymeric scaffold made up of several distinct filamentous proteins of varying lengths and stiffnesses,  which helps the cell maintain its shape and provides it support.  Most previous biophysical studies of cytoskeletal networks have focused on purified gels or networks consisting of one type of filament  \cite{hinner,gardel,tharmann,liu,kees,head,wilhelm,emt,Onck,NatMat,Koenderink, MacKPRL95, Morse98, MacReview97, BauschNatPhys,gardelfilamin, wagnerpnas2006,karenpre, chaseprl}. The cytoskeleton, however, contains three major types of filaments: microtubules (MTs), filamentous actin (F-actin), and intermediate filaments. F-actin and intermediate filaments show thermal bending on micron length scales, and often act collectively as networks with a sub-micron mesh size. MTs, on the other hand, have a persistence length of the order of millimeters and effectively behave as rigid rods on cellular lengthscales. The mechanics of a cytoskeletal composite consisting of F-actin and MTs  can, therefore, be better understood by modeling it as a  fiber-reinforced elastic composite with  stiff rods randomly and isotropically distributed in a much softer elastic matrix, rather than in terms of a network or elastic continuum made of just one type of material. There have been many fundamental studies of  the engineering properties of composite materials  \cite{Hill1965,Hashin,Eshelby,Mori,Thorpe,Thorpe2} that attest to the highly synergistic collective material properties of these materials. However, these studies often assume components of comparable stiffnesses or fibers aligned in special directions, and hence have limited applicability to the biological composites discussed above. 

Recent studies \cite{kilfoil,Brangwynne,Rodriguez,Geraldo,Mdas} on reconstituted composite cytoskeletal networks have shown that  synergistic mechanical interaction between F-actin and microtubules indeed lead to viscoelastic properties very distinct from one-component networks \cite{hinner,gardel,tharmann,liu,kees,head,wilhelm,emt,Onck,NatMat,Koenderink, MacKPRL95, Morse98, MacReview97, BauschNatPhys,gardelfilamin, wagnerpnas2006,karenpre, chaseprl}. 
A remarkable recent experimental finding on the elasticity of composite cytoskeletal networks was the appearance of
enhanced compressibility of F-actin networks upon the addition of MTs \cite{kilfoil}.  Many materials found in nature are nearly incompressible, which implies that when under external mechanical stress or strain, they try to conserve their volume, by changing their shape. This is because they have a much larger bulk modulus than shear modulus. 
It is intriguing, therefore, that the addition of very stiff rod-like MTs would give rise to a finite compressibility for such a composite of F-actin and microtubules, while pure F-actin matrices appear to be incompressible. 
 
In this paper we report detailed investigations on the mechanical response of a composite of MTs embedded in an F-actin matrix by modeling it as a composite material consisting of rods in an elastic matrix \cite{3dcomprods}. We use a dipole approximation for the rod-like inclusions and treat the composite as an effective medium whose properties we study using both micro-mechanical and continuum approaches.  We carry out our investigations for elastic matrices in both three and two dimensions. 
We report flow diagrams for the Poisson's ratio $\nu$ with increasing rod density that exhibit a stable fixed point at $\nu=1/4$ in 3D and at $\nu=1/3$ in 2D, both of which correspond to a Cauchy solid with equal Lam\'e coefficients $\lambda$ and $\mu$. 
Thus, adding rods to 3D matrices characterized by a Poisson's ratio $1/4<\nu<1/2$ makes the material \emph{more compressible} relative to the shear compliance (i.e., $\nu$ decreases), while for materials with $\nu<1/4$, stiff rods lead to a less compressible medium. In 2D, we find enhanced compressibility for matrices characterized by Poisson's ratio $1/3<\nu<1$, and a decrease in compressibility for those with a Poisson's ratio $\nu<1/3$.  We also evaluate the effective medium elastic moduli of the composite as functions of the concentration of rod-like inclusions using a self-consistent approximation. While our self-consistent approach is only approximate at intermediate concentrations, we obtain exact results in the limits of high and low concentrations. 

The rest of this paper is organized as follows. We describe our model and methods of exploring the collective mechanics of the composite under investigation in Section \ref{model}; in \ref{continuum} we describe a continuum approach to studying a fiber-reinforced composite with a 3D matrix, while in section \ref{micromechanical}  we elucidate a micromechanical method informed by linear response theory. We discuss the self-consistent calculation and the ensuing results in Section \ref{self-consistent}. In section \ref{batchelor} we describe an approach for studying this system that takes into account the full tension profile along the fibers. In section \ref{two-d} we describe the calculations for a two dimensional matrix.  We conclude with a discussion of the implications of our findings for the cytoskeleton and other composite materials in Section \ref{discussion}.

\section{Model}
\label{model}

The main focus of our study is a composite of F-actin and MTs in the cell cytoskeleton.  As discussed earlier, MTs are three orders of magnitude stiffer than F-actin.  F-actin forms a three dimensional meshwork in the cytoskeleton, with an average mesh size of  $\sim 100 nm$.  MTs, on the other hand,  effectively behave as rigid rods and are generally  $1 - 10$ microns long in most cells, whereas in axons their length can be $50 - 100$ microns.  The simplest model that can capture the essential underlying mechanics of this composite is then a collection of isotropically distributed stiff elastic rods in a comparatively soft elastic background. We further consider the limit where there is no sliding of the MTs relative to the matrix. There is increasing experimental evidence that MTs are directly mechanically coupled to other structures in the cytoplasm, including F-actin matrices. The nature of this interaction is not fully understood and it has been proposed that these interactions are mediated by microtubule associating proteins (MAPS). In the limit that the strength of this coupling is large, and the movement of the MTs is further stearically hindered by the presence of other stiff structural elements, such as other MTs or stress fibers, the MTs can be considered rigidly embedded in their matrix. We construct an isotropic and homogeneous effective medium with the same mechanics as this fiber reinforced composite of stiff rods embedded in an elastic background. We calculate the effective medium shear modulus $\mu$ and longitudinal modulus $\lambda$ for our system, followed by other mechanical quantities as such the Poisson's ratio. We use two different methods in our study:  a macroscopic approach guided by methods in continuum mechanics, and a micro-mechanical method based on linear response theory.

\begin{figure}
\includegraphics[width=8cm]{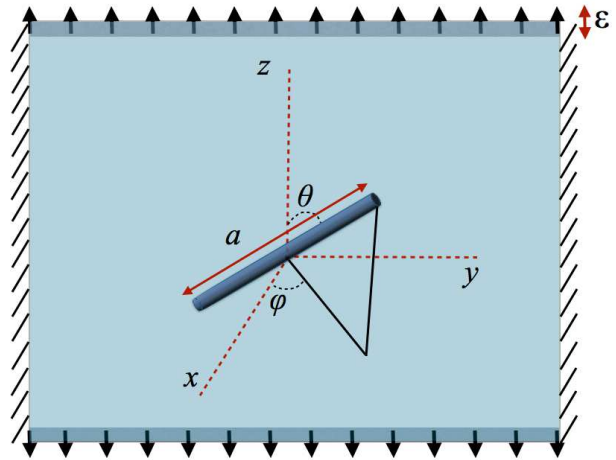}
\caption{\label{Fig1} (Color Online) We consider a representative element of the effective medium made of the background elastic matrix and rods. Here we show one representative rod oriented in a direction as shown, with a polar angle $\theta$  and azimuthal angle $\phi$.}
\end{figure}

\subsection{Continuum Approach}
\label{continuum}

For an isotropic and homogeneous elastic material under the action of external forces, the stress tensor $\sigma_{ij}$ is related to the strain tensor $u_{ij}$ by the expression
$\sigma_{ij} = \lambda \delta_{ij} u_{ii} + 2 \mu u_{ij}$, where $\mu$ and $\lambda$ are the Lam\'e constants of the material, $\delta_{ij}$ is the Kronecker Delta function and summation over repeated indices is implied. For such a material, the displacement field $u_i$ at a position $\vec r$ due to a force $\vec f$ acting at point $\vec{r'}$ is given by $u_i(\vec{r})=\alpha_{ij} (\vec{r}-\vec{r'}) f_j (\vec{r'})$,
where $\alpha_{ij}$ is the elastic response function or Green's function for the material. The response function $\alpha_{ij} (\vec{r})$ has just two distinct components, corresponding to the response parallel and perpendicular to $\vec r$: 
\bea
\alpha_{ij} (\vec{r})&&\!\!\!\!\!= \alpha_{\parallel}(r)  \hat{r}_i\hat{r}_j  + \alpha_{\perp}(r) ( \delta_{ij} - \hat{r}_i\hat{r}_j ), \nonumber\\
                             &&\!\!\!\!\!=\frac{1}{8 \pi \mu r} \big[ \hat{r}_i\hat{r}_j  (1- \beta) +  \delta_{ij} (1 + \beta)\big],
                             \label{response3d}
\eea
where
\bea
 \alpha_\parallel (r) &=& \frac{1}{4\pi\mu r} \nonumber \\
 \alpha_\perp (r) &=& \frac{(1+\beta)}{8\pi\mu r}.
 \label{alphaperpandparellel3d}
 \eea
Here, $\beta$ provides a simple measure of the degree of compressibility of the material, relative to the shear compliance. Specifically, it is given by the ratio of the shear modulus $\mu$ to the longitudinal modulus: 
\begin{equation}
\beta=\frac{\mu}{\lambda + 2\mu}.
\label{betadefinition}
\end{equation}
For incompressible materials, $\beta=0$ and the parallel and perpendicular response functions are related by a simple factor of two: $\alpha_{\parallel}(r)=2 \alpha_{\perp}(r)=1/4 \pi\mu r$. These correspond to the elastic analogue of the Oseen Tensor \cite{landau,alex} for incompressible materials in 3D. 

A related measure of a material's 
compressibility is the Poisson's ratio.  It is the ratio, when a material is stretched or compressed in one direction due to an externally applied force, of the transverse strain 
(perpendicular to the applied force), to the axial strain (in the direction of the applied force). The Poisson's ratio is defined in terms of the Lam\'e constants as $\nu= \frac{1}{2}  \lambda/(\lambda+\mu)$ in 3D and has an upper bound of $0.5$ (for incompressible materials) and a lower bound of $-1$ (corresponding to zero bulk modulus). For a composite material made up of an elastic matrix and rods, the addition of rods will lead to additional stresses in the medium and changes in the Lam\'e coefficients, compressibility and Poisson's ratio of the composite, which we calculate as follows. 

We model the composite as an isotropic and homogeneous effective medium that is made of the bare elastic medium (e.g., the F-actin matrix) and a collection of rods (MTs) embedded in it, and with elastic properties macroscopically indistinguishable from those of the actual composite.  
We consider a 3D sample of this material under uniaxial compression or extension, i.e., we apply a uniaxial strain $\epsilon_{zz}$ along the $z$ direction, with all other strain components zero. We consider a rod of length $a$ embedded in this effective medium, making polar and azimuthal angles $\theta$ and $\phi$ with respect to the coordinate axes as shown in the schematic Fig.\ \ref{Fig1}.  
When the material is strained as described above, the rod suffers a change  $\epsilon a$ in end-to-end distance, where $\epsilon=\epsilon_{zz}\cos^2\theta$. 
For the sake of simplicity, we use a dipole approximation for the constraint of fixed length $a$ of the rods that is assumed to be small compared to all other length scales in the problem. The strength of the induced dipole can be calculated as follows. 
Replacing the rod by a (tensile) point dipole of strength $p$ and orientation $\hat a$ at the center of mass of the rod, we obtain a net change in end-to-end distance, which we set to zero. This gives
\beq
0=\epsilon a+2p\alpha_\parallel'(a/2)=\epsilon a-\frac{2p}{\mu a^2},\label{DipoleStrength}
\eeq
where the prime denotes derivative. The rigid rod is thus mechanically equivalent to a dipole of strength $p=\chi\epsilon$, where 
\bea\chi=\mu\pi a^3/2.\eea
The above represents the linear response of the medium. We note that there is also a linear order change in the orientation of the rod. Given that the strength of the induced dipole is linear in the strain, we can neglect the effect of the reorientation of the rod. 
 
\begin{figure}
\includegraphics[width=8cm]{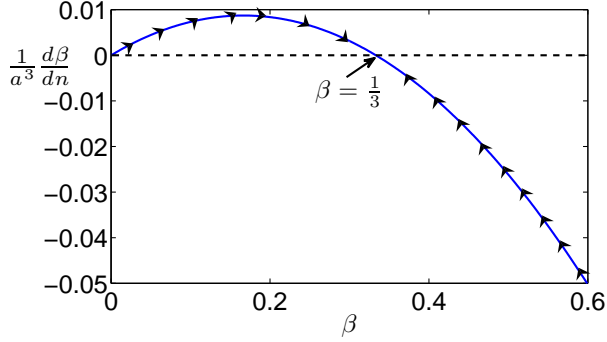}
\caption{\label{Fig2} (Color Online) The flow diagram for the degree of compressibility $\beta$ showing a stable fixed point  at $\beta = 1/3$, and an unstable fixed point at $\beta =0$.}
\end{figure}

For rod orientations in a given solid angle $d\Omega=\sin\theta d\theta d\phi$, the stress arising due to these induced dipoles is given by $\delta \sigma_{ij}= \frac{n}{4\pi} \chi\epsilon  \hat{a}_i\hat{a}_j d\Omega$.  The presence of these rods, with a number density $n$, will lead to changes in the Lam\'e constants of the material, thereby  giving rise to additional stresses at the boundary  given by $\delta \sigma_{xx} =\delta \sigma_{yy} = \delta \lambda \epsilon_{zz}$ and $\delta \sigma_{zz}= (2 \delta \mu + \delta \lambda) \epsilon_{zz}$, for an isotropic distribution of such rods. Thus, the change in the Lam\'e constants are given by:
 \begin{eqnarray}
 \delta \lambda &=& n\chi  \int \cos(\theta)^2  \sin(\theta)^2  \cos(\phi)^2 \frac{d\Omega}{4\pi} \nonumber \\
2 \delta \mu + \delta \lambda  &=&   n\chi \int \cos(\theta)^4 \frac{d\Omega}{4\pi} 
\end{eqnarray}
Solving for $\delta \mu$ and $\delta \lambda$, we find  
\beq
\delta \mu=\delta\lambda= \frac{1}{15} \chi n. 
\eeq 
Using the definition of $\beta$ and $\chi$, the above expression translates to the equation 
\beq
d\beta = \frac{\pi}{30}a^3\beta(1-3\beta) n 
\eeq
for $\beta$, the measure of compressibility relative to the shear compliance. Thus, $\beta$ increases for $0<\beta<1/3$ and decreases for $\beta>1/3$, while it remains unchanged for $\beta=0$ and $\beta=1/3$. This implies that for a medium that is only slightly compressible to begin with ($\beta$ small, but positive), adding rods makes it more compressible, while for a highly compressible medium ($\beta > 1/3$) the rods make it less so, as illustrated in Fig.~ \ref{Fig2}. This suggests a stable fixed point (to the addition of rods) at $\beta=1/3$ and hence Poisson's ratio $\nu=1/4$. Similarly, $\beta=0$ and $\nu=1/2$ corresponds to an unstable fixed point.  Our results are qualitatively consistent with recent microrheology experiments on a composite of microtubules embedded in filamentous actin \cite{kilfoil}, which reported enhanced compressibility ($\nu<0.5$) when stiff microtubules were added to an almost incompressible actin matrix ($\nu\simeq 0.5$), as inferred from the measured parallel and perpendicular response functions. 

So far we have considered the rods to be inextensible. 
%Cellular MTs have a very high but finite Young's modulus $\sim 1$ GPa. 
We now consider the rods to have finite stretch modulus $K={\cal{A}} E_r$, where ${\cal{A}}$ is the cross-sectional area of the rod with Young's modulus $E_r$. We apply an extensional strain $\epsilon$ and once again use a dipole approximation for the rod such that this dipole of strength $p=\chi\epsilon$ can cancel the end-to-end displacement of the rod caused due to the applied strain. We can write down the resulting force balance for an extension $\Delta$ of the rod.
\begin{equation}
\Delta =  a\epsilon -  \frac{2p}{\mu \pi a^2}=\frac{p}{K} \label{ForceBalance}
\end{equation}
giving,
\beq
\chi=\frac{\pi a^3 \mu}{(2+\pi a^2\mu/K)}. 
\eeq

In our effective medium theory, the Lam\'e coefficients $\lambda$ and $\mu$ for a composite with a density $n$ of rods are, therefore, 
given by the following nonlinear relations:
\begin{equation} 
\delta \lambda = \delta \mu =  \frac{\pi}{30} \frac{\mu a^3}{\left(1 + \frac{\pi a^2\mu}{2K}\right)} n.\label{diffEq}
\end{equation}
In the limit of very stiff rods, this reduces to the expression derived earlier for inextensible rods, while for highly compliant rods, this is consistent with the elastic modulus $\delta\mu=\frac{1}{15}\varphi E_r$ of an affinely deforming rod network of volume fraction $\varphi={\cal{A}} a n$. 

\begin{figure}
\includegraphics[width=8cm]{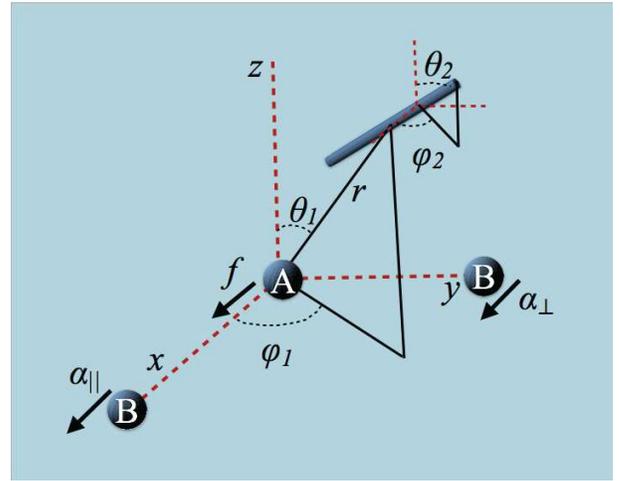}
\caption{\label{Fig3} (Color Online) We consider a point-force $f\hat x$ applied at the origin ($A$) and calculate the response at points $B$ located parallel (here, along the $x-$axis) and perpendicular (along the $y-$axis) to the applied force. The rod center of mass is located at a distance $r$ from the origin, and makes a polar angle $\theta_1$  and azimuthal angle $\phi_1$. The rod is oriented in a direction as shown.}
\end{figure}

\subsection{Micromechanical method}
\label{micromechanical}

We now investigate the micro-mechanics of our system using linear response theory that also provides the guiding principles for the micro-rheology experiments used to study the F-actin-MT composite in \cite{kilfoil}. Here, we can think of this change in the response as arising from a cloud of induced dipoles in the elastic continuum. We calculate the  displacement field $u_i$ at a position $\vec r$ in an isotropic and homogeneous elastic material due to a force $\vec f$ acting at the origin (for simplicity), using the response function $\alpha_{ij}$ in Eq.\ (\ref{response3d}).

For a composite material made up of an elastic matrix and rods, the application of a force $\vec{f}$ at, say, a point $A$ will elicit a net axial deformation of a rod located elsewhere in the medium due to its finite length. The presence of rigid rods gives rise to constraints on the displacement field induced by the applied force. The collective elastic response of the composite at a position $B$ depends on not only the applied force and the bare elasticity of the matrix, but also on the stiffness and concentration of rods. We calculate the change in the response function and Lam\'e coefficients upon addition of rods to the background elastic matrix by, once again, using a dipole approximation for the constraint of fixed length of the rods. 

We consider a single rod of length $a$ and orientation $\hat a$ embedded in the elastic medium as shown in the schematic figure \ref{Fig3}.  For simplicity, we consider an applied force at the origin of magnitude $f$ directed along the $x$-axis. The c.m. of the rod is assumed to be at $r=(x,y,z)$. We assume the rod length $a$ is small compared with $r$ and the separation between $A$ and $B$. The force $\vec{f}$ leads to a net relative displacement 
of the ends of the rod, given by
\bea
\Delta u_i=(\vec a\cdot\vec\nabla)\alpha_{ij}(\vec r)f_j=\gamma_{ijk}(\vec r)f_ja_k,
\eea
and an axial strain, given by
\bea
\epsilon=\hat a_i(\hat a\cdot\vec\nabla)\alpha_{ij}(\vec r)f_j=\hat a_i\gamma_{ijk}(\vec r)f_ja_k,
\eea
where
\bea
\gamma_{ijk}(\vec r)=\nabla_k\alpha_{ij}(\vec r).
\eea
Here, we have kept only the leading terms in $a$. Also, we approximate the constraint of the rod by an induced dipole at its c.m. of strength $p=\chi\epsilon=\chi\hat a_i\hat a_k\gamma_{ijk}(\vec r)f_j$.
This gives rise to a displacement at a point $\vec b$ (representing one of the points $B$ in Fig.\ \ref{Fig3}):
\bea
\delta u_i(\vec b)=p\hat a_j\hat a_k\gamma_{ijk}(\vec b-\vec r).
\eea
This defines a correction
\bea
\delta\alpha_{ij}=\chi\gamma_{ikl}(\vec b-\vec r)\gamma_{jmn}(\vec r)\hat a_k\hat a_l\hat a_m\hat a_n
\eea
to the response function. 
For a uniform concentration $n$ of isotropically distributed rods, the correction to the parallel component is given by 
\bea
\delta\alpha_{\parallel}(b)&=&n\chi\iint \frac{d\Omega d\Omega' d^3r}{4\pi}\gamma_{xkl}(b\hat x-\vec r)\gamma_{xmn}(\vec r)\hat a_k\hat a_l\hat a_m\hat a_n\nonumber\\
&=&n\int_0^\infty\langle\delta\alpha_\parallel(b,\rho)\rangle\rho^2\; d\rho,
\eea
where $\rho=r/b$ and
\bea
\langle\delta\alpha_\parallel(b,\rho)\rangle=\chi b^3&&\!\!\!\!\!\!\!\!\iint \frac{d\Omega d\Omega'}{4\pi}\\
&&\gamma_{xkl}(b[\hat x-\hat\rho])\gamma_{xmn}(b\hat\rho)\hat a_k\hat a_l\hat a_m\hat a_n\nonumber
\eea
represents an average over the orientation $\Omega$ of $\vec a$ and an integral over the orientation $\Omega'$ of $\vec r$. 
Similarly, by evaluating the displacement field at $\vec b=b\hat y$, we obtain
$\delta\alpha_{\perp}(b)$. 

In order to evaluate $\delta\alpha_{\parallel,\perp}$, we perform both a full Taylor series expansion of 
$\langle\delta\alpha_{\parallel,\perp}(b,\rho)\rangle$ for small $\rho$ (i.e., $\rho < 1$), as well as an asymptotic expansion for large $\rho$ (i.e., $\rho >1$).
Following the angular integrals above, we are left with just three non-zero terms, in $O(\rho)$, $O(1/\rho^2)$ and $O(1/\rho^4)$, giving the following simplified expressions for the change in the components of the response function.
\begin{widetext}
\bea
\rho^2 \langle  \delta \alpha_{\parallel} (b,\rho) \rangle
  &=& \frac{\pi a^3}{{450 b^4 \mu}}\times
    \begin{cases}   
        2(-3 + 2 \beta + 11 \beta^2)\rho  
        \qquad\qquad\qquad\qquad\qquad\qquad\quad\;\;  \text{if  $\rho<1$} \\  
        (-15 +10\beta-20\beta^2)/\rho^2 + (9  - 36 \beta + 27\beta^2 )/\rho^4  \qquad  \text{if $\rho>1,$} \end{cases} \\ \nonumber
\rho^2 \langle  \delta \alpha_{\perp} (b,\rho) \rangle
  &=& \frac{\pi a^3}{{450 b^4 \mu}}\times
    \begin{cases}  
        ( 3 - 2 \beta - 11 \beta^2)\rho 
        \qquad\qquad\qquad\qquad\qquad\qquad\quad\quad\;\;\; \text{if  $\rho<1$} \\  \nonumber
        (-15  + 10\beta - 20\beta^2)/\rho^2  + (18 - 27 \beta  + 9 \beta^2 )/\rho^4  \qquad \text{if $\rho>1$.} \end{cases}
\eea
\end{widetext}
The remaining integrals over $\rho$ lead to
\bea
\delta \alpha_{\parallel} &=& -(\pi/30)n a^3\alpha_\parallel \nonumber \\
\delta \alpha_{\perp} &=& \frac{1}{2}(1+3\beta^2)\delta\alpha_\parallel. 
\label{changeinalphaperpandparallel3d}
\eea
Expressing $\alpha_{\parallel}$ and $\alpha_{\perp}$ in terms of $\mu$ and $\lambda$ using Eq.\ (\ref{betadefinition}) and 
Eq.\ (\ref{alphaperpandparellel3d}), we obtain  $\delta \mu=\delta\lambda= \frac{1}{15} \chi n$ describing the change in the Lam\'e coefficients of the effective medium with rod density, as in the continuum approach. 

%\begin{figure}
%\includegraphics[width=8cm]{alphaparalleloct262010.eps}
%\includegraphics[width=8cm]{alphaperpoct262010.eps}
%\caption{\label{Fig4} (Color Online) The top and bottom panels respectively show the  the parallel and perpendicular components  $\rho^2 \langle  \delta \alpha_{\parallel,\perp} (\rho) \rangle_{\theta,\phi}$ as a function of compressibility measure $\beta$ and distance $\rho$, for shear modulus of the matrix $\mu=1$. The lengthscales  $a=1$ and $b=1$.}
%\end{figure} 

\section{Self consistent calculation}
\label{self-consistent}

\begin{figure}
\includegraphics[width=9.5cm]{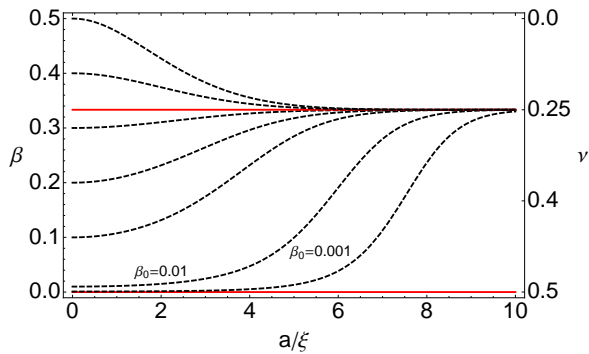}
\caption{\label{Fig5} The  degree of compressibility $\beta$ and Poisson's ratio $\nu$  of the composite as a function of mesh size $\xi$ for inextensible rods, for different values of the bare degree of compressibility $\beta_0$ of the medium. The solid (red) lines correspond to the stable and unstable fixed points at $\beta=1/3$, $\nu=1/4$ and $\beta=0$,
$\nu=1/2$ respectively.  The mesh size $\xi$ is related to the rod density $n$ by $1/\xi^{2}\equiv n a$.}
\end{figure}

We now use a self-consistent approximation wherein each added rod sees the composite as an isotropic and homogenous effective medium, with the Lam\'e coefficients $\lambda$ and $\mu$ described by their effective medium values. This is similar to self-consistent methods employed for aligned fiber-reinforced composites \cite{Hill1965}. For a small increase in the number density of rods $dn=n$, the change in the compressibility $\beta$, discussed in section II can now be described by the differential equation $d\beta/dn = \frac{\pi}{30}a^3\beta(1-3\beta)$. This suggests a way of calculating the compressibility measure $\beta$ and the Poisson's ratio $\nu$ for composites with finite rod density. The resultant $\beta$ and $\nu$ are shown in Fig.~ \ref{Fig5} for several different initial values $\beta_0$ in the absence of added rods. As earlier, we find that an incompressible material will stay incompressible (with Poisson's ratio $\nu=1/2$) even on adding rods to it, but for a matrix with  finite compressibility, however small, addition of rods tends to drive the system towards a state with $\beta=1/3$ and hence $\nu=1/4$. This suggests a stable fixed point (to the addition of rods) at $\beta=1/3$ and $\nu=1/4$, and an unstable fixed point at $\beta=0$ and $\nu=1/2$.  

\begin{figure}
\includegraphics[width=8.5cm]{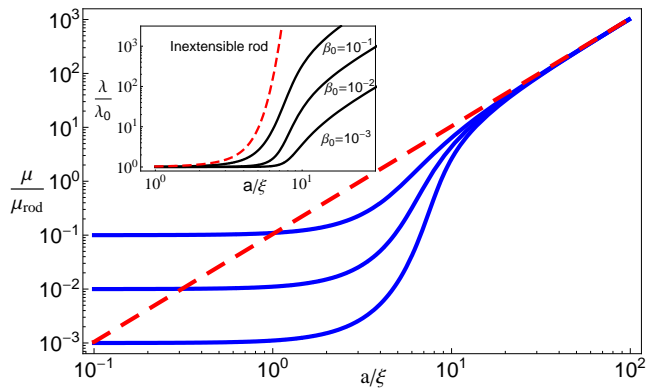}
\caption{\label{Fig6} (Color Online) The solid blue lines show the shear modulus of the composite as a function of the mesh size for different values of the ratio of the rod and medium compliance, while the dashed red line represents the affine result. The inset shows the Lam\'e coefficient $\lambda$ for different values of the initial degree of compressibility $\beta_0$ of the medium for both extensible and inextensible rods. The ratio $\mu_0/\mu_{r}=0.001$ in the inset, except for the inextensible rod (dashed line), where $\beta_0=0.1$.} 
\end{figure}

Further, this approximation allows us to cast Eq.\ (\ref{diffEq}) as a set of differential equations representing the increase of the moduli upon the addition of stiff rods.  While this represents an uncontrolled approximation and the result in Eq.\ (\ref{diffEq}) is valid at small densities where the shear modulus $\mu$ on the right-hand-side can be approximated by that of the (bare) matrix, we find that integrating Eq.\ (\ref{diffEq}) yields an exact expression in the limit of high density of the rods. The solution for $\mu$ is given by
\begin{equation}
\mu = \mu_r W \left( \frac{\mu_0}{\mu_r} \exp\left[ \frac{\mu_0}{\mu_r} + \frac{\pi n a^3}{30} \right]  \right)
\end{equation}
where,  $\mu_0$ is the shear modulus of the medium in the absence of rods and $\mu_r=2K/(\pi a^2)$. Here, $W(z)$ is the principal value of the Lambert W-function, which is defined by $z=We^W$. From this, we can also obtain the longitudinal modulus $\lambda$ as $\lambda=\lambda_0+\mu-\mu_0$ once $\lambda_0$ and $\mu_0$ for the bare background matrix are known. The results are shown in Fig.~\ref{Fig6} for various initial conditions $\mu_0$ and $\lambda_0$. For small densities $n$ and large $K$ (such that  $\mu_0/\mu_r \ll \pi na^3/3$), this reduces to $\mu\simeq\mu_0(1+\pi na^3/30)$, since $W(z)\simeq z$ for small $z$. This is consistent with the above results for dilute systems with inextensible rods. As the density of rods and corresponding shear modulus increase, however, $W(z)\simeq \ln(z)$. For highly compliant rods with $\mu_r \approx \mu_0$, this is consistent with the elastic moduli of an affinely deforming rod network of volume fraction $\varphi$: $\delta\mu=\frac{1}{15}\varphi E_r$. 
Thus the addition of rods can significantly alter the collective elasticity of the composite, and irrespective of whether the background elastic matrix is only slightly or considerably more compliant than the rods, we find that the composite stiffens significantly with added rods, finally approaching the elasticity of stiff affine rod networks at sufficiently high rod densities.

Such a self-consistent approximation as employed here is only valid in the limit of strong direct interactions between the rods. It is analogous to the self-consistent homogenization approach introduced by Hill \cite{Hill1965} and further extended in differential effective medium theories \cite{DiffEMT}. The latter model two phase composites by  incrementally adding inclusions of one phase until the desired proportion of constituents is reached.  In the limit of  weak/indirect interactions between the rods mediated by the surrounding matrix, such a self consistent approximation may no longer be valid as each rod will then have a surrounding boundary layer (at length scales smaller than its length) where it probes the bare elasticity of the matrix and not of the composite.

\section{Calculation for the full tension profile along rods}
\label{batchelor}

\begin{figure}
\includegraphics[width=8.5cm]{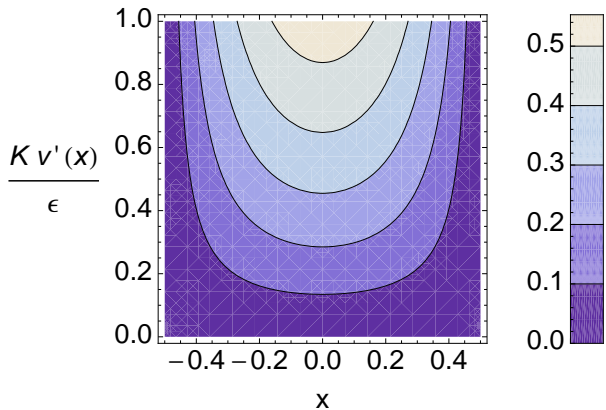}
\caption{\label{Fig7} (Color Online) The contours of the tension $K v'(x)$ divided by $\epsilon$ along a rod of length $a=1$, as a function of distance $x$ from the cm of the rod and the ratio $\mu/\mu_r$.} 
\end{figure}

In our calculations so far, we have used a  dipole approximation for the constraint on the rod. For an elastic rod of  finite length, however, the displacement field varies more smoothly than for a dipole,  \emph{decreasing} as one goes from the end of the rod toward its center. Concomitantly  the strain along the rod is uniform at its center and vanishes at its ends, as shown in Fig.~\ref{Fig7}. Our approximation of the extensional resistance of a rod by a dipole is  thus expected to overestimate the contribution of the rod to the effective medium shear modulus. One can correctly account for the tension profile along the rod using an approach analogous to
slender body theory in fluid dynamics \cite{batchelor} as follows.

Let  $v(x)$ be the displacement field along the rod in the presence of a background strain $\epsilon$ of the matrix. The tension along the rod is given by $Kv'(x)$, and the gradient of this corresponds to a net force per unit length on the rod.  This force is proportional to the relative displacement of that section of the rod with respect  to the background medium. The displacement field can be obtained from the resulting condition for force balance:
\begin{equation}
Kv''(x)=\zeta\left[v(x)-\epsilon x\right],
\end{equation}
where $\zeta$ represents the elastic coupling of the rod to the matrix. It can be thought of as the drag coefficient per unit length of the rod, similar to the viscous drag on a slender body in the presence of a background velocity field \cite{batchelor}.
For an elastic medium, we approximate $\zeta=2\pi\mu/\ln(\xi/c)$ where $c$ is the rod cross-sectional radius, as in Ref.\ \cite{Cates}. Here, the screening length $\xi$ is of order the average separation or mesh size of the rod network, which varies with rod density. However, since $\zeta$ only has a weak logarithmic dependence on $\xi$, we will treat it as a constant.  Solving the above differential equation for $v(x)$ using the condition of vanishing tension at the boundaries of the rod, $Kv'( a/2)=Kv'(-a/2)=0$, we have: 
\beq
v(x)= \epsilon \left[ x - \ell_0 \sech\left(\frac{a}{2 \ell_0}\right) \sinh\left(\frac{x}{\ell_0}\right) \right].
\eeq
Here, $\ell_0=\sqrt{K/\zeta}$ represents the length over which the longitudinal state of strain of the 
rod varies \cite{Cates}. The tension profile along the rod is shown in Fig.~\ref{Fig7}. Once again, we can represent  
the rod as a force dipole on scales large compared with the rod length $a$, but with the dipole strength  $p=\chi\epsilon$ now 
given by  $p=\int_{0}^{a/2}  2 K v''(x) x \; dx$, and therefore with $\chi$ as:
\begin{equation}
\chi=K a \left[1-  2\ell_0\tanh\left(a/2\ell_0\right)/a\right].\label{Cates}
\end{equation} 

\begin{figure}
\includegraphics[width=8.5cm]{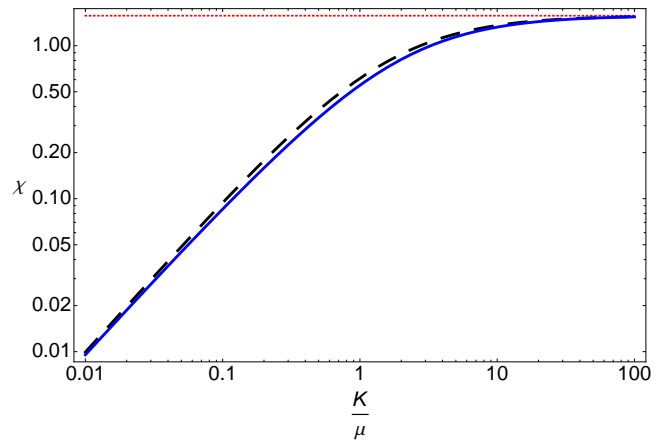}
\caption{\label{Fig8} (Color Online) The dipole susceptibility $\chi$ as a function of $K/\mu$ for infinitely stiff rods (red dotted line), compliant rods with the dipole approximation (black dashed line) and compliant rods in the Batchelor-like calculation (blue solid line) with $\mu=1$, $a=1$ and the factor $\ln \left(\xi/c \right)$ set to $1/3$.} 
\end{figure}

In the limit of highly compliant rods (small $K$), the lengthscale $\ell_0$ becomes a small length, and hence the strain and tension along the rod  is nearly constant, except very close to the ends. For very stiff rods, on the other hand, the strain exhibits a quadratic dependence, reaching a maximum at the center of the rod and vanishing at the ends of the rod. In this case, $\chi=\zeta a^3/12$, which is, as expected, somewhat smaller than the value above for the simple dipole approximation for inextensible rods. Specifically, it is smaller by a factor of $3\ln\left(\xi/c \right)$.  For intermediate values of rod compliance, i.e., if one retains the $O(1/{\ell_0}^4)$ term in the expansion of $\chi$ given by Eq.\ (\ref{Cates}) for small $1/\ell_0$, we obtain $\chi=\zeta a^3 (1- \zeta a^2/10 K)/12$. Thus, although the resulting differential equation for $\mu$ as a function of $n$ is much more complicated if one takes into account the full tension profile along the rods, we note that, apart from the prefactor of $3\ln \left(\xi/c \right)$ discussed above, the dipole strength derived in Eq.\ (\ref{ForceBalance}) takes on exactly the same limiting values for stiff and compliant rods as in Eq.\ (\ref{Cates}), and approximates intermediate values to within no more than 13\%. Figure~\ref{Fig8} shows the comparison between $\chi$ calculated above (Eq.\ (\ref{Cates})) and for the case of inextensible and compliant rods calculated using the dipole approximation, as a function of rod stiffness. Furthermore, for rods that interact directly with each other, the elastic moduli of the composite can be obtained using a self-consistent approximation \cite{Hill1965} as discussed in the previous section, while for rods that only interact with each other through their matrix, the rod concentration only enters the calculation via the screening length $\xi$. Thus, the functional forms in Figs.~ \ref{Fig5} and \ref{Fig6} are expected to be good approximations. We find that both Lam\'e coefficients once again evolve in the same way upon the addition of rods: $d\mu=d\lambda=n\chi/15$ \cite{footnote2}. This means that the qualitative form of $d\beta/dn$ in Fig.~\ref{Fig2}, as well as our conclusions regarding the fixed points at $\beta=0$ and $\beta=1/3$, remain unchanged. 

\section{Rods in a 2D elastic matrix}
\label{two-d}

We now consider a planar random fiber composite, such as rods embedded in a two dimensional membrane. Examples of planar random fiber composites include ordinary paper, synthetic and biological polymer mats and buckypaper \cite{Hall}, the last one being a prime example of a planar composite that can bear large loads.  We  apply the methods described in the previous sections to study the elastic properties of such composites. Following the continuum method described earlier, we consider a 2D effective medium representing a planar rod-based composite in the $x-y$ plane subject to uniaxial strain $\epsilon_{yy}$ along the $y$ direction, with the constraint of zero strain along the $x$ axis. We consider a single rod oriented at an angle $\theta$ with respect to the $y$-axis.  Let $\delta \lambda$ and $\delta \mu$ be the change in the Lam\'e constants due to the addition of such rods, with density $n$ and distributed isotropically, giving rise to extra stresses at the boundaries $\delta \sigma_{xx} = \delta \lambda \epsilon_{yy}$ and $\delta \sigma_{yy}= (2 \delta \mu + \delta \lambda) \epsilon_{yy}$. We use a dipole approximation for the constraint of fixed length of the rods as earlier. The rigid rod is then mechanically equivalent to a force dipole of strength $p=\chi\epsilon$,  where $\epsilon=\epsilon_{yy}\cos^2\theta$ and $\chi$ depends on the linear response of the material. For rod orientations in a given angular range $d\theta$, the stress arising from these induced dipoles is given by $\delta \sigma_{ij}= \frac{n}{2\pi} \chi\epsilon  \hat{a}_i\hat{a}_j d\theta$.  Thus,
 \begin{eqnarray}
 \delta \lambda &=& \frac{n\chi}{2\pi}  \int \cos(\theta)^2  \sin(\theta)^2   d\theta \nonumber \\
(2 \delta \mu + \delta \lambda)  &=&   \frac{n\chi}{2\pi} \int \cos(\theta)^4  d\theta 
\end{eqnarray}
Solving for $\delta \mu$ and $\delta \lambda$, we find  $\delta \mu=\delta\lambda= \frac{1}{8} \chi n$.  

\begin{figure}
\includegraphics[width=8.5cm]{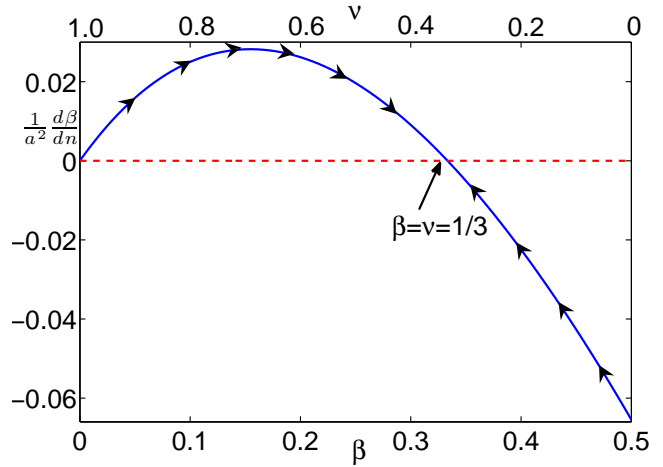}
\caption{\label{Fig9} The flow diagram for the degree of compressibility $\beta$ and Poisson ratio $\nu$ showing a stable fixed point  at $\beta = \nu= 1/3$, and an unstable fixed point at $\beta =0$, $\nu=1$.} 
\end{figure}

We calculate $\chi$ as in Eq.\ (\ref{DipoleStrength}). Here, the response function for an isotropic, homogeneous and compressible two dimensional system can also be written in terms of parallel and perpendicular components as  $\alpha_{ij} (\vec{r})= \alpha_{\parallel}  \hat{r}_i\hat{r}_j  + \alpha_{\perp} ( \delta_{ij} - \hat{r}_i\hat{r}_j )$, with the components given by  \cite{Fred-Alex-2002}: 
%\begin{widetext}
\bea
&&\alpha_{\parallel,\perp}(r)=\frac{1}{4 \pi \mu} \times\\
&&\left[ -  (1+ \beta) (\ln[r/\Lambda]+\gamma_E) + \ln[2/\beta^\beta]  \mp\frac{(\beta-1)}{2}  \right],  \nonumber
\eea
%\end{widetext}
where $\gamma_E$ is the Euler-Mascheroni constant.  The above equations describe the elastic response of the material for distances $r$ smaller than $\Lambda$, which represents an upper cut-off length, below which the elasticity of the material can be considered two-dimensional. 
Solving Eq.\ (\ref{DipoleStrength}), we find
\beq
\chi = \mu \pi a^2/(1+\beta).
\eeq 
Together with the expressions for $\delta \lambda$ and $\delta \mu$ obtained above, we find
the following differential equation for $\beta$:
\beq
 \frac{d \beta}{d n}=  \frac{\pi a^2 \left(1- 3\beta \right) \beta}{8\left(1+\beta\right)}. 
 \eeq
The Poisson's ratio $\nu$, which in  two dimensions  is defined as $\nu=\lambda/(\lambda+2\mu)$, and $\beta$ are related by the simple expression $\nu = 1 - 2 \beta$. Using the same framework as earlier, we find that at high rod densities the composite would now approach a steady state with a compressibility $\beta=1/3$ (as earlier), but Poisson's ratio $\nu=1/3$ as seen in Fig.~\ref{Fig9}. The differential equations for the Lam\'e coefficients $\lambda$ and $\mu$ are given by :
\beq
 \frac{d \mu}{d n} =  \frac{d \lambda}{d n} =  \frac{\mu \pi a^2}{8} \frac{\lambda + 2 \mu}{\lambda + 3 \mu}. \nonumber
\eeq

The elastic response in a two dimensional material has an upper cut-off distance, described in our case by $\Lambda$. For a 2D monolayer lying on a 3D viscous fluid phase, this lengthscale is set by the ratio of the two dimensional shear modulus of the monolayer to the three dimensional modulus of the fluid phase. In this case $\Lambda$ represents a crossover length below which strains are two-dimensional, and above which they are dominated by viscous damping in the three-dimensional fluid underneath the monolayer. 

\section{Discussion}
\label{discussion}

We have studied the collective mechanical response of composites of rods embedded in an elastic medium such as MTs in F-actin \cite{kilfoil,yi-chia} or carbon nanotubes \cite{nanotube} in synthetic or biological gels, using a mean field approach and a dipole approximation for constraints on the rods.  We find that  an initially incompressible material, both in three and two dimensions, will stay incompressible ($\beta=0$) even on adding rods to it, however, if the medium is even marginally compressible to begin with, the addition of rods will drive it to a fixed point at $\beta=1/3$, signifying compressibility. In three dimensions this corresponds to a stable fixed point in the Poisson's ratio at $\nu=1/4$ (compressible) and an unstable fixed point at $\nu=1/2$ (incompressible), while in two dimensions it corresponds to $\nu=1/3$ and $\nu=1$ respectively. Our results may help to explain recent experiments \cite{kilfoil} that have reported $\nu< 1/2$ for composites of microtubules and F-actin networks (in 3D). We also derive an expression for the Lam\'e coefficients as a function of arbitrary rod density. Although approximate, this calculation recovers the expected results for very low and high rod densities. Our results suggest an important role for stiff filaments  such as MTs and stress fibers in the mechanics of the cell cytoskeleton---they not only enhance the stiffness of the cytoskeleton ~\cite{yi-chia} and its ability to bear large forces \cite{Brangwynne,Mdas}, but may also endow it with a compressibility that enables it to undergo small volume changes when necessary. 

The ability to tune the Poisson's ratio of a composite by varying rod concentration has important applications for engineering materials. The primary  mechanism of failure in composite materials is through a tensile failure caused by the reinforcing fibers getting narrower and pulling away from the matrix when stretched. Allowing the composite to have a finite compressibility will lead to comparatively less thinning and stretching of the fibers for a given load, and consequently the load required to cause structural failure will significantly increase. Our result that the addition of elastic rods or fibers leads to a monotonic evolution of Poisson's ratio toward the value $1/4$  in 3D, either from above or below, is further consistent with recent numerical calculations for fiber-reinforced concrete, showing a weak increase in $\nu$ with fiber density \cite{PasaDutra}. Although our work was primarily motivated by intracellular networks that are nearly incompressible, concrete represents an interesting case of a matrix with PoissonÕs ratio of  $\sim 0.2$. While the authors of Ref.  \cite{PasaDutra} do not make any general predictions or statements concerning PoissonÕs ratio, they report specific values for certain fiber volume fractions. Interestingly, they find that fiber inclusions lead to an increasing PoissonÕs ratio ($<1/2$) for all systems studied, consistent with our general predictions in Fig.~ \ref{Fig2} and Fig.~\ref{Fig5}.  Flow diagrams in the {\em area} Poisson's ratio have been previously reported for composites made of a background isotropic matrix and unidirectional fibers arranged randomly or in a superlattice, for the plane perpendicular to the long axis of the fibers  ~\cite{Thorpe,Thorpe2}.  These studies are applicable to  2D composites of circular disc-like inclusions, 
and report a fixed point in the {\em area} Poisson's ratio $\nu=1/3$ for random and Kagome arrangement of discs.

Fiber reinforced composites, such as the system we study, can have both direct interactions between the inclusions, as well as indirect interactions through the surrounding matrix. In this manuscript we consider the limit of strong direct interactions between the fibers, and assume the fibers to be rigidly embedded in the matrix and employ a self-consistent approach in calculating the macroscopic elasticity of the composite. Our approach is similar in concept to homogenization methods \cite{Hill1965} and differential effective medium theories \cite{DiffEMT} that take the point of view that a composite material may be constructed by making infinitesimal changes in an already existing composite. 
Further investigations are needed to take into account the effect of tuning the strength of interactions between fibers, as well as of fiber curvature due to the bending elasticity of the fibers and thermal fluctuations. It is likely that fiber curvature will lead to smaller effective elastic modulus in longitudinal compression than in tension in the same direction with important consequences for the effective medium normal stresses. Furthermore, in the present study the distribution of fibers has been assumed to be homogeneous and isotropic. In biological composites, however, one often comes across highly inhomogeneous regions of fiber reinforcement, such as a cytoskeletal composite that shows bundle formation and orientational ordering of filaments. These features have important implications for the nonlinear mechanical response of the cytoskeleton and will be addressed in future work. 

We thank M Kilfoil, ME Cates, HNW Lekkerkerker, M.F. Thorpe, DC Morse, TC Lubensky and PD Olmsted for helpful discussions. MD was supported by a VENI fellowship from the Netherlands Organization for Scientific Research (NWO). FM was supported in part by FOM/NWO.


\begin{thebibliography}{}




\frenchspacing
\bibitem{alberts}  B.\ Alberts, {\it et al.},
{\it Molecular Biology of the Cell}, Fourth edition. Garland Science, NewYork, (2005).

\bibitem{howard} J. Howard, Mechanics of Motor Proteins and the Cytoskeleton, Sinauer, 2001.

\bibitem{nanotube1}  L.\ Hu {\it et. al.} Proc.\ Nat.\ Acad.\ Sci.\ {\bf 106}, 21490 (2009);
R.A.\ MacDonald {\it et. al.}, J.\  Biomed.\ Mat.\ Res.\ Part A {\bf 74 A}, 489 (2005); 
R.\ Andrews, M.C.\ Weisenberger, Curr.\ Opin.\ in Solid State \& Mat.\ Sci.\ {\bf 8}, 31 (2004).

\bibitem{MacKPRL95}F.C.\ MacKintosh, J.\ K\"as and P.A.\ Janmey, Phys.\ Rev.\ Lett.\ {\bf 75}, 4425 (1995).

\bibitem{hinner} B.\ Hinner, {\it et al.}, Phys.\ Rev.\ Lett.\ {\bf 81}, 2614 (1998).

\bibitem{Morse98} D.C.\ Morse, Macromolecules {\bf 31}, 7030-7043 (1998).

\bibitem{gardel} M.L.\ Gardel et al.,  Science 304, 1301 (2004).

\bibitem{kees} C.\ Storm, {\it et al.}, Nature {\bf 435}, 191 (2005).

\bibitem{Koenderink} G.H.\ Koenderink, {\it et al.}, Phys.\ Rev.\ Lett.\ {\bf 96}, 138307 (2006).

\bibitem{tharmann} R.\ Tharmann, M.M.A.E.\ Claessens and A.R.\ Bausch, Phys.\ Rev.\ Lett.\  {\bf 98}, 088103 (2007).

\bibitem{liu} J. Liu, {\it et al.}, Phys.\ Rev.\ Lett.\ {\bf 98}, 198304 (2007).

\bibitem{NatMat} P.A.\ Janmey, {\it et al.}, Nature Materials {\bf 6}, 48 (2007).

\bibitem{head} D.A.\ Head, A.J.\ Levine, F.C.\ MacKintosh, Phys.\ Rev.\ E {\bf 68}, 061907 (2003); Phys.\ Rev.\
Lett.\ {\bf 91}, 108102 (2003).

\bibitem{wilhelm} J.\ Wilhelm and E.\ Frey, Phys.\ Rev.\ Lett.\ {\bf 91}, 108103 (2003).

\bibitem{Onck} P.R. Onck, {\it et al.}, Phys.\ Rev.\ Lett.\ {\bf 95}, 178102 (2005).

\bibitem{emt} M.\ Das, F.C.\ MacKintosh, and A.J.\ Levine, Phys.\ Rev.\ Lett.\ {\bf  99}, 038101 (2007).

\bibitem{MacReview97}F.C.\ MacKintosh and P.A.\ Janmey, Current Opinion in Solid State \& Materials Science {\bf 2}, 350 (1997).

\bibitem{BauschNatPhys}A.R.\ Bausch and K.\ Kroy, Nature Physics \textbf{2}, 231 (2006).

\bibitem{gardelfilamin} M.L.\ Gardel, {\it et al.} Proc.\ Nat.\ Acad.\ Sci.\ {\bf 103}, 1762 (2006).  

\bibitem{wagnerpnas2006} B.\ Wagner, {\it et al.}, Proc.\ Nat.\ Acad.\ Sci.\ {\bf 103}, 38, 13974 (2006).

\bibitem{chaseprl} C. P.\ Broedersz, C.\ Storm, and F. C.\ MacKintosh, Phys.\ Rev.\ Lett.\ {\bf 101}, 118103 (2008).

\bibitem{karenpre} K. E.\ Kasza, {\it et al.} Phys.\ Rev.\ E {\bf 79}, 041928 (2009).

\bibitem{Hill1965} R.\ Hill, J.\ Mech.\ Phys.\ Solids {\bf 13}, 189 (1965).

\bibitem{Hashin}  Z.\ Hashin and B.\ W.\ Rosen, J. \ Appl.\ Mech. {\bf 29}, 143, 1962 ; J. \ Appl.\ Mech.\ {\bf 31}, 223 (1964).

\bibitem{Eshelby}  J. D.\ Eshelby, Proc.\ Royal\ Soc.\ London: Series A. Math. and Phys. Sci. {\bf 241}, 376 (1957); 
Proc.\ Royal\ Soc.\ London: Series A. Math. and Phys. Sci. {\bf 252}, 561 (1959).

\bibitem{Mori} T. \ Mori and K.\ Tanaka, Acta Metal.\ {\bf 21}, 571 (1973).

\bibitem{Thorpe}  I.\ Jaisuk, J.\ Chen and M.F.\ Thorpe, J.\ Mech.\ Phys.\ Solids {\bf 40}, 373 (1992); L.C.\ Davis, K.C.\ Hass, J. \ Chen
and M.F\ Thorpe, Appl.\ Mech.\ Rev.\ {\bf 47} S5 (1994).

\bibitem{Thorpe2} J.\ Chen, M.F.\ Thorpe and L.C.\ Davis, J.\ Appl.\ Phys. {\bf 77}, 4349 (1995).
 

\bibitem{kilfoil} V.\ Pelletier, N.\ Gal, P.\ Fournier, and M.L.\ Kilfoil, Phys.\ Rev.\ Lett.\ {\bf 102}, 188303 (2009).

\bibitem{Brangwynne} C.P.\ Brangwynne et al.,  J.\ Cell\ Bio.\ {\bf 173}, 733 (2006). 

\bibitem{Rodriguez}  O.C.\ Rodriguez et al.,  Nat.\ Cell\ Biol.\ {\bf 5}, 599 (2003). 

\bibitem{Geraldo}  S.\ Geraldo and P.R.\ Gordon-Weeks,  J.\ Cell\ Sci.\ {\bf 122}, 3595 (2009).

\bibitem{Mdas} M.\ Das, A.J.\ Levine and F.C.\ MacKintosh, Europhys.\ Lett.\ {\bf 84}, 18003 (2008).

\bibitem{3dcomprods} M.\ Das and F.C.\ MacKintosh, Phys.\ Rev.\ Lett.\ {\bf 105}, 138102 (2010).


\bibitem{landau} L. D.\ Landau, and E.M.\ Lifshitz, {\it Theory of Elasticity}, Pergamon Press, Oxford, (1986).

\bibitem{alex} A.J.\ Levine and T.C.\ Lubensky,   Phys.\ Rev.\ Lett.\ {\bf 85}, 1774 (2000).

\bibitem{DiffEMT} A. N.\ Norris, Mech.\ Mater.\ {\bf 4}, 1 (1985);  M.\ Avellaneda,  Commun.\ Pure\ Appl.\ Math {\bf 40}, 527 (1987).

\bibitem{batchelor} G. K.\ Batchelor, J.\ Fluid\ Mech.\ {\bf 46}, 813 (1971).

\bibitem{Cates} M.E.\ Cates and S.F.\ Edwards, Proc.\ Royal.\ Soc.\ of London. Series A {\bf 395}, 89 (1984).

\bibitem{footnote2}{Although Ref.\ \cite{Cates} doesn't give an explicit expression for $\lambda$, the result $d\lambda=d\mu$ is equivalent to the result in equations (4.22- 4.23) derived therein.}

\bibitem{Hall} L.\ Hall, et al. Science {\bf 320}, 504 (2008).

\bibitem{Fred-Alex-2002} A.J.\ Levine and F.C\. MacKintosh, Phys.\ Rev.\ E {\bf 66}, 061606 (2002).

\bibitem{yi-chia} Y.C.\ Lin, G.H.\ Koenderink, F.C.\ MacKintosh and D.A.\ Weitz, Soft Matter {\bf 7}, 902 (2011).

\bibitem{nanotube} P. M.\ Ajayan  and  J.M.\ Tour, Nature {\bf 447}, 1066 (2007).

\bibitem{PasaDutra} V.F.\ Pasa Dutra, S.\ Maghous, A.\ Campos Filho and A.R.\ Pachero, Cement and Concrete Res.  {\bf 40}, 460 (2010).




\end{thebibliography}
\end{document}